\documentclass{article}

\usepackage[final]{neurips_2025}

\usepackage[utf8]{inputenc} 
\usepackage[T1]{fontenc}    
\usepackage{hyperref}       
\usepackage{url}            
\usepackage{booktabs}       
\usepackage{amsfonts}       
\usepackage{xcolor} 
\usepackage{todonotes}     
\usepackage{subcaption}     
\usepackage{nicefrac}       
\usepackage{microtype}      
\usepackage{xcolor}   
\usepackage{amsmath}
\usepackage{graphicx}
\usepackage{algorithm,algpseudocode}
\usepackage{array}
\usepackage{amsthm}
\usepackage{listings}
\usepackage[english]{babel}

\usepackage{enumitem}
\usepackage{tikz}
\usepackage{multirow}

\usepackage{latexsym}
\usepackage{amssymb}

\usepackage{pifont}
\usepackage{svg}
\usepackage{pgfkeys}
\usepackage{pgfplots}
\usepackage{chemformula}
\usepackage{tikz}
\usetikzlibrary{shapes,arrows,positioning,arrows.meta,calc}

\title{A Chemically Grounded Evaluation Framework for Generative Models in Materials Discovery}

\author{%
  Elohan Veillon$^{1,2}$, Astrid Klipfel$^{1}$, Adlane Sayede$^{2}$, Zied Bouraoui$^{1}$\\
  $^{1}$Univ. Artois, UMR 8188, Centre de Recherche en Informatique de Lens (CRIL), France.\\
  $^{2}$ Univ. Artois, UMR 8181, Unitée de Catalyse et de Chimie du Solide (UCCS),  France.\\
  \texttt{\{elohan.veillon,astrid.klipfel,adlane.sayede,zied.bouraoui\}@univ-artois.fr} \\
}

\begin{document}

\maketitle

\begin{abstract}
Generative models hold great promise for accelerating materials discovery, but their evaluation often overlooks the chemical validity and stability requirements crucial to real-world applications. Density Functional Theory (DFT) simulations are the gold standard for evaluating such properties but are computationally intensive and inaccessible to non-experts. We propose a chemically grounded, user-friendly evaluation framework that integrates DFT-based stability analysis with commonly used machine learning (ML) metrics. Through systematic experiments using both perturbative and generative methods, we demonstrate that conventional ML metrics can misrepresent chemical feasibility. To address this, we propose new insights on robust metrics and highlight the importance of simulation-informed evaluation for developing reliable generative models in materials science.
\end{abstract}

\section{Introduction}
Generative models for materials discovery have garnered significant attention in recent years, offering a promising avenue for designing novel materials with desired chemical properties. 
Advances in generative diffusion models and graph-based neural networks have proven to be a powerful tool in both organic chemistry \cite{gasteiger_gemnet_2021} and materials science \cite{Chen2022}, enabling efficient exploration of chemical space. Methods such as molecular dynamics \cite{dynamol_1,dynamol_2,dynamol_3} and generative models \cite{ml_screening,PGCGM,deepmind_GNoME}, particularly diffusion models \cite{cdvae,unimat,ms-mattergen,gemsdiff}, have been widely adopted to accelerate materials generation. 
Diffusion models  typically employ powerful GNN architectures, which have proven their relevance for handling chemical data due to their ability to process graph-structured information effectively. 
However, despite these advancements, evaluating the chemical validity of ML-generated materials remains a major challenge. Current evaluation methodologies often fail to address the practical needs of chemists, leading to a misalignment between ML-based assessments and real-world feasibility.

A crucial issue lies in the choice of evaluation metrics. Many studies rely on latent- or property-based metrics, which provide insights into structural diversity or statistical similarity to training data, but fail to capture essential constraints such as stability and physical feasibility. 
In contrast, simulation-based metrics, widely used in chemistry, provide a more reliable alternative to evaluate materials. For instance,  Ab-Initio Random Structure Searching (AIRSS) \cite{airss}, which perturbs known materials and validates their stability using Density Functional Theory (DFT) calculations \cite{DFT_1,DFT_2}. DFT is regarded as the gold standard for material stability assessments due to its foundation in quantum mechanics. However, these simulations are computationally expensive, require specialized expertise, and introduce additional challenges related to scalability and reproducibility. 
Previous works on generative model evaluation \cite{Baird2024} rarely include direct comparisons between metrics based on DFT calculations and latent-based/property-based metrics. 

To bridge this gap between ML-generated structures and real-world chemical validation, we propose a streamlined, easy-to-use framework that integrates DFT simulations for stability assessments and relaxations. Our framework systematically evaluates generative models by combining DFT-based assessments with existing non-simulation-based metrics, ensuring that generated materials meet chemical feasibility requirements.
Our framework includes automation of key tasks such as structure preparation, symmetry analysis, parameter tuning, and post-processing. This reduces the need for domain expertise and makes the pipeline accessible to non-specialists. Furthermore, we enable users to control the simulation workload by selectively applying DFT to a subset of generated structures. This allows for efficient evaluation under limited computational budgets, making our framework both rigorous and scalable.
For experimental validation, we compare some state-of-the-art ML metrics with physically grounded DFT-based evaluations, demonstrating that existing ML metrics often misrepresent true chemical feasibility. Our findings highlight some shortcomings in current generative model evaluation practices, motivating the need for a new class of evaluation metrics that align with the realities of materials science. By providing a rigorous and accessible evaluation framework, this work lays the foundation for more reliable, chemistry-aware generative modeling, ultimately improving the quality and applicability of ML-generated materials.  Our goal is not only to benchmark existing metrics but also to reveal their limitations and guide the development of more realistic evaluation practices for generative materials modeling. 

The remainder of this paper is structured as follows. Section \ref{sec:background} discusses current state-of-the-art diffusion models and metrics to evaluate material generation addressed in this work and highlights their limitations. Section \ref{sec:pipeline} presents our proposed framework, outlining its key components and methodology. Section \ref{sec:experiments} describes the experimental setup and evaluates generative models using different metrics. Finally, Section \ref{sec:conclusion} concludes the paper and outlines future research directions.

\section{Evaluation Metrics for Generative Models}
\label{sec:background}

Diffusion models have recently gained significant popularity in materials science. However, evaluating such generative models demands rigorous methods that accurately assess their chemical feasibility. Traditionally, three main categories of metrics are used: simulation-based, property-based, and latent-based metrics. While latent-based and property-based metrics have been widely adopted in ML research, their relevance to real-world chemistry is often limited. These approaches primarily assess generated structures based on statistical similarity to training data, which does not ensure stability, synthesizability, or physical validity.
In contrast, simulation-based metrics offer a physically grounded evaluation of generated materials. Among these, Density Functional Theory (DFT) calculations are considered the gold standard for assessing stability and materials properties. However, their computational cost and domain expertise requirements limit accessibility. As a result, many generative models are evaluated using non-simulation-based metrics, which prioritize ease of computation over chemical realism, potentially leading to misleading evaluations. This section provides a description of widely adopted diffusion models as well as an overview of these three metric categories, discussing their strengths and limitations in the context of generative model evaluation.

\subsection{Diffusion models}

Diffusion models are a generative AI framework based on denoising architectures. They use Langevin dynamics in the sampling process to produce high-quality data. In recent years, diffusion models have entered materials science, surpassing traditional approaches like variational autoencoders and GANs. The first notable application in material science was CDVAE \cite{cdvae}, a diffusion model built upon a VAE using GemNet \cite{gasteiger_gemnet_2021}, a GNN for molecular interaction prediction. Later works \cite{gemsdiff,diffcsp} extended this idea, combining diffusion models with GNNs and similar evaluation methods.
Typically, these models are evaluated using property-based and latent-space methods, which are cost-effective but limit result quality and interoperability. MatterGen \cite{ms-mattergen} introduced a new evaluation approach using DFT calculations. Though conceptually similar to earlier models, it relies on a larger dataset and more rigorous evaluation. However, the high computational cost of DFT on thousands of structures could restrict adoption, limiting advanced generative model development to well-funded institutions.

\subsection{Simulation-based metrics}
When evaluating simulations, it is essential to define key concepts: total energy, stability, and DFT relaxation (Figure \ref{fig:simulation_metrics} in Appendix). The total energy of a material is the sum of the internal energies of its atoms and the interactions between them. Since structures are represented as periodic units in 3D space, their energy is often expressed as the average energy per atom. The stability of a material is determined by its total energy and is crucial to assess whether a chemical structure can exist and persist in nature. A lower energy per atom indicates greater stability (see the SUN metric below). Finally, DFT relaxation refers to the process of optimizing atomic positions within a structure to minimize internal forces until a local energy minimum is reached. The relaxed structure serves as a reference to compare atomic positions with the initial configuration (see the RMSD metric below).
Because DFT-based metrics such as SUN and RMSD directly evaluate how closely generated structures align with stable, real-world configurations, they are particularly intuitive for chemists.

\noindent\textbf{Stability, Uniqueness, Novelty (SUN)}
The SUN metric \cite{ms-mattergen} aims to determine the percentage of structures that are Stable, Unique, and Novel. Comparing if two structures are identical relies on the \textit{StructureMatcher} methods from the \textit{pymatgen} Python library. This structure's matcher algorithm uses a fixed threshold to check if the geometry of both structures is similar enough to consider both structures equivalent \cite{cdvae,diffcsp,ms-mattergen}. Novel indicates that the structures generated do not exist in the training dataset. Unique means that the structures generated by the model are different from each other. Finally, the stability is assessed by comparison between tested structures energy per atom and already known structures of the same chemical composition energies. In practice, we represent materials on a phase diagram that plots formation energies against chemical compositions using the convex hull method \cite{convex_hull_method}. The convex hull of the phase diagram corresponds to the lowest-energy minima known for the corresponding composition. When we refer to stability, we look at the energy "above the hull" to denote the distance between the hull and the structure. A structure is considered stable if it is close enough to the convex hull. 
Figure \ref{fig:convex_hull} in Appendix shows an example of convex hull, where all known relaxed structures are represented by black dots. The x-axis represents the \ch{Ti} to \ch{O} ratio and the y-axis the formation energy, so columns of dots represent different conformations of the same structures with different energy levels. The green band over the convex hull is the region below the arbitrary threshold, inside which structures are considered stable. All structures that are not in this region but still below the direct line between elemental references are called "metastable" structures. Once all known structures are plotted on the diagram and the convex hull is drawn, we plot the tested structure and see if it is in the "stable" region. If it is, it is considered stable for the SUN metric.

\noindent\textbf{Root Mean Squared Displacements (RMSD).} The RMSD is used to determine the similarity between generated and DFT relaxed versions of a structure. This metric calculates the average distance between atoms in the two structures. When the structures are more alike, the RMSD value is smaller. Its value is always positive, and decreases toward zero when structures are more alike. Note that due to periodic boundary conditions, the distance d between two atoms is the length of the shortest path from one position in the 3D torus to the other.
\begin{equation}
    \text{RMSD}(x^\text{gen},x^\text{DFT}) =\sqrt{\frac{1}{N}\sum^N_{i=1}d(x^\text{gen}_i,x^\text{DFT}_i)^2}
\end{equation}
The final global reported RMSD is the average of the RMSD of each structure. Calculating RMSD requires careful consideration of a few key factors. Two chemical structures may be identical, but their atoms may not be in the same order. Because crystals can be identical through permutation of atom order, translation and/or rotation of the unit cell, RMSD calculation has to ensure that compared crystals are properly aligned with respect to these criteria to minimize the metric value.

\noindent\textbf{Limitations.} Because of their computational intensiveness, simulation metrics (particularly RMSD) can only be computed for a limited number of generated structures in a reasonable time. Additionally, DFT simulations typically depend on proprietary software, like VASP \cite{VASP_1,VASP_2}, which needs a paid licence and specific theoretical knowledge to use properly. 

\subsection{Property-based metrics}\label{sec:prop_based_metrics}
Another approach to evaluate generative models is to compare specific properties between generated and test set structures. The quality of the generated data is then defined as the similarity between the generated data and the test set data. 
This is illustrated in Figure \ref{fig:property_based_metrics} in the Appendix where a property of structures from the generated set and the test set can be estimated and then compared with the earth mover's distance (EMD) metric.
If most of the structures in the training data are stable, then a good generator is one that can produce stable structures. This concept can be applied to all material properties such as density, band gap, minimum interatomic distance, among others. Evaluating these types of metrics can be done much faster than simulation-based metrics if the properties can be calculated directly from structure parameters (e.g., the density of the structures) or assessed using ML models (e.g., the formation energy of the structures). 

\noindent\textbf{Earth mover's distance (EMD).} 
Also known as the Wasserstein-1 Distance, the EMD quantifies the difference between two probability distributions. It is commonly used to compare the properties of generated structures with those of a test set. Initially, they have been introduced to compare density and formation energy \cite{cdvae}. To apply this, a specific property is calculated for all structures in both the generated and test sets. These properties are treated as samples from underlying probability distributions. The EMD serves as a measure of distance between these distributions. Consequently, metrics to evaluate generative models can be defined by calculating the EMD for a given property between the set of generated structures and the set of test structures.

\noindent\textbf{Validity}
This metric has been introduced in \cite{Court2020} and identifies physically invalid structures by checking for atom pairs with distances shorter than 0.5 Å (50 pm), as such configurations are unrealistic. The validity score is computed as the percentage of valid structures in the dataset.

\noindent\textbf{Limitations.} The accuracy of property estimation can be questionable. Most regression models show very impressive performance on various prediction tasks. However, regression models are usually trained on stable materials, and the accuracy of a regression model on unstable structures or out-of-distribution data is uncertain. The choice of properties may also be questionable. Indeed, some properties are extremely important and deserve to be similar between generated and test structures, such as formation energy. However, the relevance or absence of other properties is not entirely clear. This observation may limit the confidence we can place in metrics based on property distributions.

\subsection{Latent-based metrics}

Latent-based metrics are conceptually similar to property-based metrics, but instead of comparing a specific property between generated and test data, they compare the distributions of latent vectors. This can be achieved by using a feature generator to produce a fingerprint of the structures or by extracting latent vectors from a pre-trained model without its final layers. There are two main approaches to compare these latent vector distributions. As illustrated in Figure \ref{fig:latent_metrics} in the Appendix, this process parallels property-based metrics. However, instead of evaluating a one-dimensional property of the materials, an abstract latent vector is generated, allowing for the comparison of multi-dimensional probability distributions using more advanced methods.

\noindent\textbf{Coverage.} The coverage metrics \cite{cdvae} uses CrystalNN to generate fingerprints \cite{crystalnn_fingerprints} to compare two sets of materials. It uses precision and recall calculations to compute the overlap between the generated structures with the test structures. In fact, we can define a good generative model as one that produces structures in the same domain as the dataset used. In this context, existing work has defined accuracy as the percentage of generated structures that are in the domain of the dataset. Alternatively, a good generative model must also be able to generate structures in the entire domain of its dataset. Recall is an estimate of the percentage of the training domain that can be generated by the generative model. The domain definition of the generated data is similar to the domain of the dataset if and only if precision and recall are good.

\noindent\textbf{Fréchet Distance.} Another approach is to use a similarity measure to evaluate whether the probability distribution of the generated data and the test set are similar. Previous work has generally used a Fréchet distance to compare the latent vector distribution \cite{NIPS2017_8a1d6947}. This means that the comparisons between the generated structures and the structures of the test set are not compared against their domains but against their probability distributions.

\noindent\textbf{Limitations.} These approaches capture information about the generative capability of ML models. However, they also lack interpretability. The coverage metrics and the Fréchet distance are difficult to interpret in terms of physical laws. As a result, a given value of the aforementioned metrics is difficult to relate to the actual relevance of a generative model for material design.
%


\section{Evaluation Framework for Generative Models}
\label{sec:pipeline}


To bridge the gap between machine learning-based material generation and chemically grounded evaluation, we propose a unified framework that integrates simulation-based and non-simulation-based metrics. This pipeline is designed to (i) ensure chemical feasibility, (ii) automate complex simulation steps, and (iii) allow comparative evaluation across different generative strategies.
\footnote{Our dataset and implementation are available at \url{https://github.com/E-Veillon/metrics_pipeline}}.

\subsection{Overview}
The framework is designed to be modular and scalable, enabling researchers to evaluate the chemical feasibility of generated materials while maintaining accessibility for non-experts. The evaluation process consists of three core stages. First, candidate materials are generated using two distinct strategies (Section \ref{sec:generation}): a perturbative approach, where known materials are modified through controlled distortions, and a generative approach, where new structures are synthesized using diffusion-based machine learning models. This dual approach allows us to analyze metric behavior under both synthetic perturbation and realistic generation scenarios. Second, the generated structures undergo simulation and screening via DFT (Section \ref{dft-sec}). This stage automates key steps such as parameter tuning, structure relaxation, and property extraction, reducing the manual burden typically associated with quantum simulations. We use symmetry analysis, duplicate filtering, and element-based heuristics to ensure physical plausibility before simulation. Notice that, although DFT is traditionally resource-intensive, we design our framework to automate its use and make simulation-based validation both accessible and scalable.
Finally, we compute a range of evaluation metrics across all candidates (Section \ref{ch:stability_band_gap}). These include physically grounded metrics derived from DFT (e.g., RMSD, energy-above-hull for SUN), as well as faster, approximate metrics (e.g., EMD, latent-space coverage). This allows for both in-depth and lightweight assessments depending on the computational budget.
Our pipeline is fully automated, compatible with large-scale datasets such as the Materials Project (MP) and OQMD, and is intended to serve as a robust benchmark for evaluating generative models in materials science. Figure \ref{fig:pipeline} illustrates the full pipeline and its modular stages.


\tikzstyle{block} = [draw, node distance=2.1cm,fill=white, rectangle, minimum height=3em, text width=6em, align=center, rounded corners=.2cm]
\tikzstyle{output} = [fill=white, node distance=2cm, text width=5em]
\begin{figure*}
    \centering
    \scriptsize
    \begin{tikzpicture}[auto, >=latex']
        \node [block] (dataset) {Dataset};
        \node [block,below of=dataset, node distance=1cm] (diffusion) {Generate candidates};
        \node [block,right of=diffusion] (preprocess) {Pre-processing};
        \node [block,right of=preprocess] (relax) {DFT};
        \node [block,right of=relax] (screening) {Stable\\ screening};
        \node [block,below of=screening, node distance=1.4cm] (metrics) {Metrics\\ calculation};
        \node [output,right of=metrics] (outmetrics) {metrics};
        \node [block,right of=screening] (bandgap) {$\Delta$-sol};
        \node [output,right of=bandgap] (output) {structures};
        
        \draw[-> ] (dataset) -- (diffusion);
        \draw[-> ] (diffusion) -- (preprocess);
        \draw[-> ] (preprocess) -- (relax);
        \draw[-> ] (relax) -- (screening);
        \draw[-> ] (screening) -- (bandgap);
        \draw[-> ] (bandgap) -- (output);
        
        \draw[rounded corners=0.3cm,-> ] (diffusion) |- (metrics);
        \draw[-> ] (screening) -- (metrics);
        \draw[-> ] (metrics) -- (outmetrics);
    \end{tikzpicture}
\caption{Our framework has three main components: (1) “generation,” including preprocessing; (2) “simulation,” involving two DFT-based simulations; and (3) “metrics calculation".}
\label{fig:pipeline}
\end{figure*}
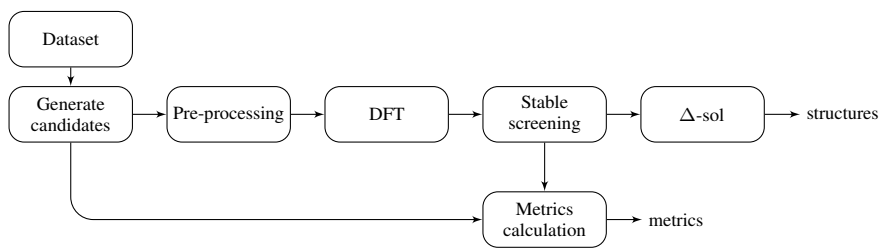

\subsection{Generation}
\label{sec:generation}
This step aims to generate potential new materials to evaluate using our screening pipeline. We propose two different methods using random structures generated from known materials (i.e. perturbative approach) and generation based on diffusion models (i.e. generative model approach). By using multiple diffusion models, we will be able to compare their performance on different metrics and under real conditions. This study is provided in Section \ref{sec:experiments}.
%
We consider Materials Project (MP)\footnote{\url{https://next-gen.materialsproject.org/}} \cite{Materials_project} and OQMD\footnote{\url{https://oqmd.org/}}\cite{OQMD_1,OQMD_2} as datasets that contain a large number of structures obtained using DFT relaxation. The MP contains more than 150.000 structures, while OQMD contains more than 1.2 million structures. These datasets allow evaluation of metrics under realistic conditions with commonly used chemical compositions.
Existing work in the literature has typically used small subsets of datasets containing only small structures of up to 24 atoms \cite{cdvae}. In this work, while we also use datasets generated from DFT simulations,  we perform the pre-processing without any size limit and focus on large general databases. To do this, we filter out MP and OQMD data to keep meta-stable structures with a formation energy of less than 0.1 eV/atom but with no restriction on the size of the structures. For diffusion models with a limit on the size of the generated structures, we limited the possible number of atoms to 64. The final datasets contain up to 87k structures for MP and up to 221k structures for OQMD.

\noindent\textbf{Perturbative approach.} To better understand the behaviour of the different metrics proposed in Section \ref{sec:background}, introduce three strategies of  perturbations: (1) a perturbation mode in which the atomic positions are randomly modified within a maximum distance, (ii) a perturbation mode in which the lattice parameters are modified, i.e. the angles and lengths of the lattice, and (iii) a perturbation mode combines the perturbations to atomic positions and lattice parameters. We have also defined three levels of perturbation: small, medium and large. For small perturbations, the atomic positions are moved by a maximum of 0.1 angstroms and the lattice parameters are moved by 1 degree and a maximum of 0.1 angstroms. For medium perturbations, the atomic positions are shifted by a maximum of 0.5 angstroms and the lattice parameters are shifted by a maximum of 3 degrees and 0.5 angstroms. For large perturbations, the atomic positions are shifted by 1.5 angstroms, and the lattice parameters are shifted by 5 degrees and 1.5 angstroms. The parameters used for the perturbations are summarised in the Table \ref{tab:perturbative_parameters} in Appendix.

\noindent\textbf{Generative models approach}
We also used three generative material models: CDVAE \cite{cdvae}, GemsDiff \cite{gemsdiff} and DiffCSP \cite{diffcsp}. The models are trained using the original code provided in their original papers. We also used the configuration of the models provided in their original papers. To allow a fair comparison, we did not use a pre-trained model and retrained all models with the pre-processed datasets as described above. SUN metrics are computed over 1024 structures with static DFT  and RMSD calculation on 64 structures due to the higher computational cost of relaxation.

\subsection{DFT simulation}
\label{dft-sec}
The framework is designed to automate DFT simulations with minimal user intervention. It handles DFT parameters tuning and metrics calculation automatically, making it accessible even to users without expertise in chemistry or computational materials science. The DFT parameters in our framework are derived from a well-established DFT calculations setup used in the Materials Project database to assess material properties (see Appendix \ref{ch:vasp_setup} for more details).
We now describe how to perform geometry optimization. This simulation aims to identify the nearest thermodynamically stable configuration of atoms and the lattice to the original structure.

\noindent\textbf{Post-processing} When it is done, the structures can be compared before and after optimization to determine the generative model's capability of generating stable or metastable structures. 
We use the \emph{SpacegroupAnalyzer} of pymatgen as an interface to the spglib library \cite{spglib} to find the space symmetry group of structures and evaluate how well generative models can create structures with internal symmetry. To remove duplicate structures before the simulation we used \emph{StructureMatcher} from pymatgen, which helps group similar structures based on a defined tolerance. We used a tolerance of 20\% for lattice lengths, 30\% for atomic coordinates, and 5 degrees for lattice angles.


\noindent\textbf{Filtering out unlike structures} 
The periodic table consists of various elements with different properties. Some elements, like the noble gases, which are present in the last column of the table, are relatively unreactive and are unlikely to be found in any real material. On the other hand, the elements in Block F, which are present in the two rows separated from the rest of the table, have specific properties that make them unsuitable for all applications. These elements require a more complex simulation than what is currently implemented to be accurately described, especially when considering spin-orbit coupling. The method used to estimate band gaps does not support these elements (see Section \ref{ch:stability_band_gap}). Lastly, structures with atoms placed too closely together are not realistic, as electrostatic repulsion would prevent such arrangements. It is advisable to remove such structures. One can adjust the minimum distance that all atoms must maintain from each other in a structure, but it is not recommended to set a value lower than the default of 0.5 angstroms, which is approximately equal to the radius of a hydrogen atom to only keep valid structures as defined in Section \ref{sec:prop_based_metrics}.


\subsection{Stability and Band gap estimation}\label{ch:stability_band_gap}

We now explain how to evaluate meta-stability based on the total energy calculated using DFT.

\noindent\textbf{Phase diagrams construction}
We retrieve known reference structures (database or training dataset) that correspond to the compositions (different chemical elements, e.g. Ti-O diagram means all structures with Ti and O atoms only) of the generated materials to compare them with what already exists. We construct the diagram with the reference structures, then we compare generated structures against the fixed diagram to compare their energy to the reference. Notice that larger compositions (i.e., those with more elements) can include smaller compositions (e.g., Fe-Ti-O includes all Ti-O, Fe-Ti, and Fe-O structures). We first calculate the smaller diagrams and then, if necessary, include the smaller structures in the larger diagrams.

\noindent\textbf{Computing Energy with convex hull}
After placing all the structures on the diagram, the energy above the hull is calculated for each generated structure. This is done by finding the difference between the energy of the structure and that of the hull, at the same compositional coordinate. The stability of the structures is then measured based on their energy above the hull. The structures are said to be strictly stable if their energy above hull is inferior or equal to 0.0 eV/atom. The S.U.N. metric proposed by Zeni et al. \cite{ms-mattergen} has a tolerance of 0.1 eV/atom above the hull to consider a structure as stable, as some experimentally observed crystals are known to lie slightly above the hull. Hence, we follow their work and define the default stability threshold at 0.1 eV/atom.

\noindent\textbf{Band Gap}
We use the $\Delta$-Sol method \cite{delta-sol} to calculate the band gap. This method for determining band gaps is relatively low in computational cost, thus allowing for the screening of many structures. However, since this method is semi-empirical (with $N^*$ fitted to a set of results obtained for selected structures), it can provide inaccurate values in some cases (e.g., band gaps greater than 4 eV) compared to other more general methods such as the Modified Becke-Johnson (MBJ) potential \cite{MBJ_method}, or a hybrid functional such as  HSE06 \cite{HSE_functional}. Therefore, it incorporates the ability to calculate uncertainty for the obtained value by reproducing the same calculation protocol as for direct determination, but for minimum and maximum tabulated values of $N^*$\cite{delta-sol}. 


\section{Evaluation and Analysis}
\label{sec:experiments}
This section provides a deep analysis of the different metrics on data generated using either a perturbative approach or diffusion models. We give a   insights into the evaluation process, offering suggestions for the development of new generative models for material science. 

\begin{table*}[t]
\centering
\scriptsize
\begin{tabular}{ccc|c|ccc|ccc}
\toprule
        \multirow{2}{*}{Dataset} & \multirow{2}{*}{Perturb.} & \multirow{2}{*}{Scale} & \multicolumn{1}{c|}{Simulation} & \multicolumn{3}{c|}{Properties} & \multicolumn{3}{c}{Latent space} \\
        & & & RMSD $\downarrow$ & Val. $\uparrow$ & EMD($\rho$) $\downarrow$ & EMD($E$) $\downarrow$ & Prec. $\uparrow$ & Rec. $\uparrow$ & FD $\downarrow$ \\
        \toprule
         MP & lattice
        & small & 0.0393 & 100.00\% & 0.0149 & 0.0042 & 99.98\% & 100.00\% & 0.0107 \\
        & & medium & 0.1814 & 100.00\% & 0.0758 & 0.0214 & 99.95\% & 99.93\% & 0.0776 \\
        & & large & 0.5185 & 70.89\% & 0.4836 & 0.1393 & 99.40\% & 98.67\% & 0.8538 \\
         & positions
        & small & 0.0718 & 100.00\% & 0.0085 & 0.0038 & 99.98\% & 100.00\% & 0.0205 \\
        & & medium & 0.4294 & 99.26\% & 0.0085 & 0.1092 & 100.00\% & 98.52\% & 0.9471\\
        & & large & 1.1472 & 87.83\% & 0.0085 & 0.5177 & 99.95\% & 76.95\% & 6.7702\\
         & lat. + pos.
        & small & 0.0690 & 100.00\% & 0.0167 & 0.0047 & 100.00\% & 100.00\% & 0.0249\\
        & & medium & 0.4732 & 99.52\% & 0.0615 & 0.1252 & 99.98\% & 94.12\% & 1.0812\\
        & & large & 1.3372 & 59.65\% & 0.4826 & 0.5720 & 99.84\% & 59.95\% & 7.4880\\
         \midrule
         OQMD & lattice
        & small & 0.0491 & 100.00\% & 0.0414 & 0.0088 & 99.86\% & 99.95\% & 0.0488 \\
        & & medium & 0.1642 & 100.00\% & 0.1341 & 0.0282 & 99.86\% & 99.86\% & 0.0951 \\
        & & large & 0.5261 & 100.00\% & 0.8185 & 0.1259 & 99.29\% & 99.62\% & 0.5554 \\
         & positions
        & small & 0.0591 & 100.00\% & 0.0321 & 0.0094 & 99.86\% & 99.93\% & 0.0524 \\
        & & medium & 0.2848 & 99.95\% & 0.0321 & 0.0531 & 99.95\% & 98.60\% & 0.3205 \\
        & & large & 1.2030 & 98.00\% & 0.0321 & 0.2367 & 99.88\% & 74.88\% & 1.9096 \\
         & lat. + pos.
        & small & 0.0764 & 100.00\% & 0.0326 & 0.0102 & 99.90\% & 99.93\% & 0.0546 \\
        & & medium & 0.3340 & 99.90\% & 0.1374 & 0.0682 & 99.93\% & 95.45\% & 0.3826 \\
        & & large & 1.1366 & 97.69\% & 0.8029 & 0.2832 & 99.83\% & 57.86\% & 2.2868 \\
         \bottomrule
    \end{tabular}
    \caption{This table presents the effect of controlled perturbations (lattice, atomic positions, and combined) on various evaluation metrics across different datasets (MP and OQMD). Metrics include RMSD, SUN, EMD on both the density and formation energy, and latent space similarity measures (precision, recall, and Fréchet distance). The scale of perturbation is reported in Table \ref{tab:perturbative_parameters}.}
    \label{tab:perturbation}
\end{table*}

\begin{table*}[t]
\centering
\scriptsize
\begin{tabular}{cc|cc|ccc|ccc}
\toprule
        \multirow{2}{*}{Dataset} & \multirow{2}{*}{Model} & \multicolumn{2}{c|}{Simulation} & \multicolumn{3}{c|}{Properties} & \multicolumn{3}{c}{Latent space} \\
        & & RMSD $\downarrow$ & SUN $\uparrow$ & Val. $\uparrow$ & EMD($\rho$) $\downarrow$ & EMD($E$) $\downarrow$ & Prec. $\uparrow$ & Rec. $\uparrow$ & FD $\downarrow$ \\
        \toprule
         MP & CDVAE
        & \textbf{0.3084} & 0.0\% & \textbf{100.0\%} & 0.4580 & 0.1649 & 97.46\% & 97.17\% & 1.7679 \\
         & GemsDiff
        & 0.3976 & 0.99\% & 99.12\% & \textbf{0.1330} & \textbf{0.0528} & 97.46\% & 97.46\% & \textbf{0.7030} \\
         & DiffCSP
        & 0.3141 & \textbf{2.25\%} & 95.12\% & 0.2882 & 0.3346 & \textbf{98.34\%} & \textbf{97.66\%} & 2.4638 \\
         \midrule
         OQMD & CDVAE
        & 0.4287 & 0.19\% & \textbf{99.90\%} & 0.9092 & 0.2432 & 98.24\% & 90.53\% & 1.8002 \\
         & GemsDiff
         & 0.3289 & 3.81\% & 98.63\% & \textbf{0.0938} & \textbf{0.0515} & 95.80\% & 97.56\% & \textbf{0.4263} \\
         & DiffCSP
        & \textbf{0.3114} & \textbf{8.69\%} & 99.80\% & 0.1834 & 0.0608 & \textbf{99.68\%} & \textbf{99.50\%} & 0.6368 \\
         \bottomrule
    \end{tabular}
    \caption{Performance of CDVAE, GemsDiff, and DiffCSP on two datasets (MP and OQMD). The analysis includes simulation-based metrics (RMSD, SUN), property-based evaluations (EMD for density and energy), and latent-based evaluations (precision, recall, and Fréchet distance).}
    \label{tab:diffusion}
\end{table*}

\noindent\textbf{Evaluation of perturbation approach}
The goal is to compare all evaluation metrics against a well-defined dataset of perturbed materials, where increasing perturbation levels correspond to greater structural instability. By analyzing how metrics respond to these controlled perturbations, we can better understand what each metric measures and how it behaves. Since these perturbations are applied randomly, they provide a neutral benchmark, free from predefined assumptions about material stability. Table \ref{tab:perturbation} presents the results of this perturbative approach.

\noindent\textbf{Evaluation of diffusion models} To complete our metric analysis, we include structures generated by diffusion models, as described in Section \ref{sec:generation}. Our objective is to evaluate the metrics in a realistic context, observing how they perform when applied to state-of-the-art generative models. A key motivation is to determine whether the results shown in Table \ref{tab:perturbation} also hold for diffusion models. Notably, diffusion models may exhibit biases that differ from the Gaussian noise used in the perturbative approach. This comparison allows us to verify whether the insights derived from the perturbative method remain valid for structures generated by diffusion models. Table \ref{tab:diffusion} summarizes the evaluation outcomes for the diffusion-based approach.



\subsection{Quantitative results}
We now examine different metrics, which can be categorized into three distinct groups: (i) metrics such as RMSD and SUN, which are more theoretically grounded, but their implementation implies several limitations; (ii) Metrics that lack simulation support, which are considered outdated; and (iii) More dependable metrics that also do not rely on simulations. 

\noindent\textbf{RMSD and SUN.}
RMSD and SUN metrics exhibit only partial correlation. As illustrated in Table \ref{tab:diffusion}, the top-performing results for SUN do not consistently correspond to the top results for RMSD. However, Table \ref{tab:perturbation}, which details various levels of perturbation, reveals that the metrics for different generative models show similar behavior under medium-sized perturbations. This comparison offers an initial assessment of the performance of the generative models.
Generative models are known for their ability to produce novel and unique structures, as explained in \cite{ms-mattergen}. However, the true challenge lies in generating stable structures. RMSD evaluates how close a structure is to a local energy minimum, while SUN measures its proximity to the global energy minimum. As such, SUN is often considered a more valuable metric, as it represents the fraction of structures that are genuinely useful to chemists. However, SUN is sensitive to the type of DFT calculations performed and the reference database employed. The setup of the DFT calculation has an impact on the precision of the total energy \cite{cances2023density}. The type of functional and setup of the DFT calculation is chosen to be a compromise between precision and computational cost. But the energy above the hull required to be considered meta-stable is quite small compared to the precision of DFT calculation because of its high cost. As a result, inconsistency in the DFT setup between multiple work limits their comparability. This effect can be shown when we compare our SUN metric with \cite{ms-mattergen}. In principle, we should obtain the same result for CDVAE and DiffCSP, but we can only see that our numbers are off by a large margin. The most probable explanation is simply a small modification of the DFT setup that causes this gap. Notice that even with a large difference regarding the SUN metric, the RMSD metric is comparable.

\noindent\textbf{Validity, Precision and Recall.} 
Metrics such as validity, precision, and recall are not effective in distinguishing diffusion models' performances, as they tend to yield uniformly high scores regardless of the model’s performance. While these metrics might have been informative when generative models were less advanced, they are insufficient to evaluate existing models. This limitation is particularly evident under significant perturbations, as illustrated in Table \ref{tab:perturbation}. For example, even with an RMSD exceeding 1 \AA, the validity and precision metrics remain unusually high.

\noindent\textbf{Earth mover's distance and Frechet distance.} 
While these metrics offer a more nuanced comparison between models compared to other non-simulation metrics, they still have notable limitations. There appears to be some correlation between EMD, Fréchet Distance and simulation metrics, as indicated in Table \ref{tab:perturbation}. However, this correlation is less precise when comparing generative models and their generated structures. For example, models like GemsDiff exhibit significantly lower metrics compared to other generative models, suggesting that DiffCSP generally performs better. Only CDVAE consistently underperforms across all metrics used.
\emph{These findings raise questions about the efficacy of non-simulation metrics. Indeed, simulation-based metrics are deemed more physically grounded by chemists, yet the model rankings vary depending on whether simulation metrics are employed.}
Metrics like EMD offer valuable insights into the limitations of generative models. The density metric, $\rho$, is particularly noteworthy as it assesses whether the generated cells are physically meaningful. As noted in previous studies, multiple lattice representations can describe the same crystal structure; while their shapes may vary, the lattice volume and thus the density, remains invariant. Supercell constructions further alter the lattice shape and volume but preserve the density. Consequently, this metric enables consistent comparison between crystalline structures, regardless of their specific lattice representations. Our analysis shows that GemsDiff generates more geometrically plausible cells than CDVAE and DiffCSP. However, since GemsDiff structures exhibit lower stability, DiffCSP appears more effective at producing stable chemical compositions and atomic arrangements. Such nuanced insights cannot be captured by EMD alone, underscoring the complementary value of this metric.

\subsection{Discussion}

Our work introduces a practical, physics-based framework for systematically comparing generative models in materials science. While previous studies have proposed new architectures or metrics, few have examined how standard machine learning metrics align with physical feasibility. This lack of standardization leads to inconsistent results across studies, as varying metric definitions and simulation settings often produce conflicting evaluations.

Our experiments reveal that some evaluation metrics are more effective than others in capturing the physical quality of generated structures. Simulation-based metrics such as RMSD and SUN provide reliable indicators of structural stability and thermodynamic feasibility, but they should not be used in isolation. DFT relaxation, for instance, can artificially increase the number of “stable” structures since it optimizes geometries to lower total energy. Consequently, SUN values differ in meaning depending on whether they are computed from relaxed or static calculations. Because relaxation tasks are computationally expensive, this approach may favor well-funded institutions, whereas smaller teams may struggle to perform such calculations at comparable precision. Using static DFT evaluations offers a more affordable and equitable alternative for assessing stability.
Recent diffusion-based models also exhibit high novelty and uniqueness, making it valuable to jointly assess stability and diversity to better understand the realism of generated structures. Complementary metrics such as EMD and Fréchet Distance provide additional insights into how well generated samples capture the overall data distribution. Even when RMSD and SUN scores are strong, distributional metrics can reveal underrepresented regions in the reference dataset, highlighting their importance in comprehensive model evaluation.


EMD and Fréchet Distance are computationally efficient, making them well-suited for large-scale screening. Their strength lies not in replacing simulation-based metrics but in complementing them, providing a broader and faster view of generative model performance. The usefulness of EMD can be further enhanced by grounding it in physically meaningful quantities. Instead of relying on ML-predicted properties, EMD can be computed over DFT-derived energies or other simulated properties such as band gaps, paving the way for energy- or band gap–based EMD variants that combine scalability with physical relevance.
We also observe a clear difference in model performance between the MP and OQMD datasets. Diffusion models generally perform better on OQMD, though the reasons remain uncertain. Contributing factors may include OQMD’s larger size, greater chemical diversity, higher-precision DFT data, or inherent construction biases. Despite its scale and richness, OQMD remains underexplored in generative modeling. Its extensive, DFT-validated structures make it a promising resource for studying model generalization and robustness across chemical domains.


\section{Conclusion}
\label{sec:conclusion}
This paper presents a comprehensive evaluation of generative models for materials discovery, emphasizing the reliability of commonly used metrics. By comparing simulation-based and non-simulation-based approaches, we highlight their complementary strengths and limitations. Among non-simulation metrics, EMD and Fréchet Distances best capture distributional differences but fail to reflect physical feasibility, underscoring the continued importance of simulation-based metrics like RMSD and SUN. While these are more chemically grounded, they remain computationally costly. We therefore advocate a hybrid evaluation strategy, which uses lightweight metrics for screening and DFT-based assessments for high-confidence candidates, to develop a standardized, physics-aware framework for rigorous and interpretable model evaluation in materials science.

\section*{Acknowledgment}
This project has received financial support from the CNRS through the MITI interdisciplinary programs through its exploratory research program (PRIME CNRS AIM-GPT).

\bibliography{mybibfile}
\bibliographystyle{abbrv}

\newpage
\appendix
\onecolumn

\section{Evaluation Metrics}

\begin{figure}[h]
    \centering
    \includegraphics[scale=0.8]{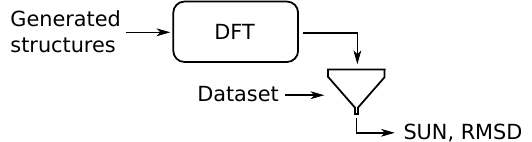}
    \caption{Simulation-based metrics workflow representation.}
    \label{fig:simulation_metrics}
\end{figure}

\begin{figure}[h]
    \centering
    \input{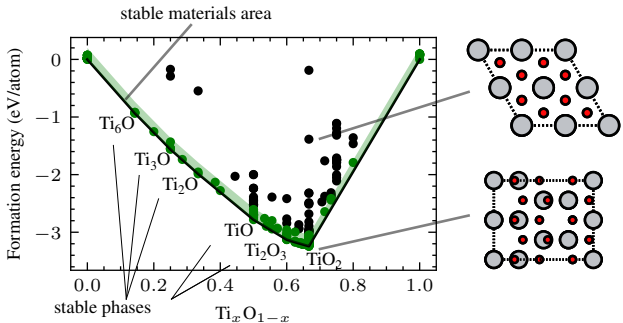}
    \caption{Phase diagram with convex hull on energy of the \ch{Ti-O} chemical system. Each dot plots a structure. Structures inside and below the green area are considered metastable when their energy level is under 0.1 eV/atom above the hull.}
    \label{fig:convex_hull}
\end{figure}

\begin{figure}[h]
    \centering
    \includegraphics[scale=0.8]{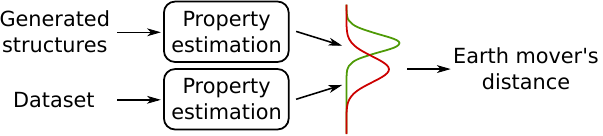}
    \caption{Property-based metrics are based on a similarity measure (generally EMD) between distributions of a given property.}
    \label{fig:property_based_metrics}
\end{figure}

\begin{figure}[h]
    \centering
    \includegraphics[scale=0.8]{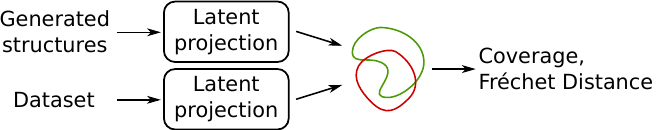}
    \caption{Latent-based metrics used a pre-trained machine learning model to generate vector embeddings of a set of structures. Then, methods for comparing vector sets, such as coverage metrics and Fréchet distance, are often employed to analyze and assess the similarities or differences between two sets of vectors.}
    \label{fig:latent_metrics}
\end{figure}

\begin{table}[h]
\footnotesize
    \centering
    \begin{tabular}{c||c|c|c|c}
    \toprule 
        metric & prop. & latent & sim. & reference \\
     \midrule
        validity (struct.) & \ding{56}& & & \cite{Court2020}\\
        density (EMD) & \ding{56} & & & \cite{cdvae}\\
        energy (EMD)$^*$ & \ding{56} & & \ding{56}
        & \cite{cdvae}\\
        recall (cov.) & & \ding{56} & & \cite{xu2021learning}\\
        precision (cov.) & & \ding{56} & & \cite{xu2021learning}\\
        Frechet distance & & \ding{56} & & \cite{gemsdiff}\\
        RMSD & & & \ding{56} & \cite{ms-mattergen}\\
        SUN & & & \ding{56} & \cite{ms-mattergen}\\
        \bottomrule
    \end{tabular}
    \caption{Category and original work of the available metrics in the framework. $^*$The category of energy EMD depends on how said energy is obtained. If it is calculated with DFT it is a simulation metric, otherwise (e.g. ML regression) it is a property metric.}
    \label{tab:metrics}
\end{table}

\section{INCAR configuration used for VASP simulations}\label{ch:vasp_setup}

The Vienna Ab-initio Simulation Package (VASP) quantum computation software \cite{VASP_1,VASP_2} is used to compute DFT level total energies and structure optimizations. The Projector Augmented Wave (PAW) formalism \cite{PAW_potentials} is used for pseudo-potentials, and the Perdew, Burke and Ernzerhof (PBE) Generalized Gradient Approximation (GGA) is used as DFT functional \cite{PBE_functional}. Note that we use a more recent version of the PBE associated pseudo-potentials than the pymatgen default, the "PBE 5.4" pseudo-potentials. 

For the static step used to obtain total energies for stability comparisons in order to compute the SUN metric, the base of the configuration for INCAR generation is the pymatgen preset \emph{MPStaticSet}. For the relaxation step used to compute the RMSD metric, the base of the configuration is the pymatgen preset \emph{MPRelaxSet}. \footnote{Both \emph{MPStaticSet} and \emph{MPRelaxSet} details can be found at \url{https://pymatgen.org/pymatgen.io.vasp.html}}. For both steps, we make few modifications to better fit our needs:

First, as we do not have any \emph{a priori} knowledge about the electronic structure of the tested materials, we use the Gaussian Smearing method instead of the default Tetrahedron method with Blöchl corrections (ISMEAR = 0 instead of -5), because it is non-specialized and more robust for a wide range of materials.

Second, we only need total energy for the static step, and initial and final structures for the relaxation step, which are both reported in main output files. Therefore, we minimize specific output files generation from VASP by removing LORBIT keyword, and setting LCHARG, LWAVE and LVHAR to "False".

Lastly, we set the GGA+U parameters to correspond to the ones fitted in the work of Jain et al. \cite{MIT_relax}.

\begin{table}[t]
    \centering
    \small
    \begin{tabular}{c|c|c|c}
        \toprule
        perturbation & atomic position & lattice angle & lattice length \\
        \midrule
        small & 0.1\AA & 1.0° & 0.1\AA \\
        medium & 0.5\AA & 3.0° & 0.5\AA \\
        large & 1.5\AA & 5.0° & 1.5\AA \\
        \bottomrule
    \end{tabular}
    \caption{Parameters of the perturbative approach}
    \label{tab:perturbative_parameters}
\end{table}
\end{document}